\documentclass[twocolumn,prl,epsfig,rotate,showpacs,cjk,superscriptaddress,english]{revtex4}
\usepackage[T1]{fontenc}
\usepackage{CJKutf8}
\usepackage{amsmath}
\usepackage{amssymb}
\usepackage{graphicx}
\usepackage{esint}

\makeatletter
\@ifundefined{textcolor}{}
{%
 \definecolor{BLACK}{gray}{0}
 \definecolor{WHITE}{gray}{1}
 \definecolor{RED}{rgb}{1,0,0}
 \definecolor{GREEN}{rgb}{0,1,0}
 \definecolor{BLUE}{rgb}{0,0,1}
 \definecolor{CYAN}{cmyk}{1,0,0,0}
 \definecolor{MAGENTA}{cmyk}{0,1,0,0}
 \definecolor{YELLOW}{cmyk}{0,0,1,0}
 }

\usepackage{dcolumn}
\usepackage{epsfig}
\usepackage{bm}
\usepackage{babel}

\makeatother

\begin{document}
\begin{CJK}{UTF8}{gkai}

\title{Effects of Discrete Breathers on Heat Conduction}

\author{Daxing Xiong(熊大兴)}
\address{Department of Physics and Institute of Theoretical Physics and Astrophysics, Xiamen University, Xiamen 361005, Fujian, People's Republic of China}
\address{Department of Physics, Fuzhou University, Fuzhou 350002, Fujian, People's Republic of China}
\author{Jiao Wang(王矫)}
\author{Yong Zhang(张勇)}
\email{yzhang75@xmu.edu.cn}
\author{Hong Zhao(赵鸿)}
\address{Department of Physics and Institute of Theoretical Physics and Astrophysics, Xiamen University, Xiamen 361005, Fujian, People's Republic of China}

\date{\today}
\begin{abstract}
Intensive studies in the past decades have suggested that the heat conductivity $\kappa$ diverges with the system size $L$ as $\kappa\sim L^{\alpha}$ in one dimensional momentum conserving nonlinear lattices and the value of $\alpha$ is universal. But in the  Fermi-Pasta-Ulam-$\beta$ lattices with next-nearest-neighbor interactions we find that $\alpha$ strongly depends on $\gamma$, the ratio of the next-nearest-neighbor coupling to the nearest-neighbor coupling. We relate the $\gamma$-dependent heat conduction to the interactions between the long-wavelength phonons and the randomly distributed discrete breathers. Our results provide an evidence to show that the nonlinear excitations affect the heat transport.
\end{abstract}

\maketitle

In the studies of non-electronic heat conduction, it is an
important progress to realize that the heat conductivity $\kappa$
diverges with the system size $L$ as $\kappa\sim L^{\alpha}$ in one dimensional momentum conserving nonlinear chains \cite {LepriReport,DharAdvPhys}. The value of the exponent $\alpha$ is believed to be universal \cite{LiviNature}, though there are lots of debates on what value(s) it takes. (e.g., if there exists one universal class with $\alpha=\frac{1}{3}$ \cite{key-14,key-8} or two with $\alpha=\frac{1}{3}$ and $\alpha=\frac{2}{5}$ \cite{key-7,key-9,key-10} is still controversial.) The universality of $\alpha$ roots in the theory first proposed by Peierls \cite{PeierlsPhonon}, where the essence of the non-electronic heat conduction at low temperatures is modeled as a weakly interacting phonon gas. Based on this model, a universal heat conduction law constrained only by the dimensionality of a system regardless of the details of its microscopic dynamics is thereby expected.

On the other hand, at high temperatures nonlinear excitations such as traveling solitary waves \cite{SolitonRev} and discrete breathers (DBs) \cite{DBRev} are ubiquitous in nonlinear lattice systems. Hence interactions between phonons and nonlinear excitations should be studied and taken into account when their effects are considerable. As nonlinear excitations involves microscopic dynamical details, if a  universal heat conduction law still exists certainly deserves careful investigations. In this respect quite few studies have been reported. Early work of our group showed that traveling solitary waves may play an important role in heat conduction of the Fermi-Pasta-Ulam-$\beta$ (FPU-$\beta$) chains \cite{Zhao2005,Zhao2006}, but this was argued against by some authors \cite{D.CaiPRL,BLiEffectPhonons}. Besides, DBs have also been proposed as a phonon scattering mechanism \cite{FlachDBscattering} for the normal heat conduction numerically observed in the harmonic chains with nonlinear on-site potentials \cite{fai4DBs} and in the rotator chains \cite{rotator}. In spite of these studies, at present whether --- and if yes how the nonlinear excitations would affect the heat transport in low dimensional momentum conserving systems is still an open question.

In this Letter we present a study to show how DBs play a role in the
heat conduction in one dimensional nonlinear lattices.  Specifically, we perform numerical analysis to investigate the heat conduction in FPU-$\beta$ chains with the next-nearest-neighbor (NNN) coupling. Our results show clearly that $\alpha$ varies continuously with the ratio of the NNN coupling to the nearest-neighbor (NN) coupling, suggesting our system does not belong to any universality class characterized by a constant $\alpha$. Moreover, we find $\alpha$ is correlated to the overlap of the phonons' spectra and the DBs' spectra, and in such a sense we relate the $\gamma$-dependent heat conduction behavior to the phonon scattering by the DBs.

Our system is a chain of $N$ identical particles with both the NN and NNN interactions \cite{key-18} whose Hamiltonian is
\begin{equation}
H=\sum_i[\frac{p_{i}^{2}}{2 \mu}+
V\left(q_{i+1}-q_{i}\right)+
\gamma V\left(q_{i+2}-q_{i}\right)].
\end{equation}
Here $q_{i}$ is the displacement of the $i$th particle from its equilibrium position and $p_{i}$ its momentum, and the potential is of the FPU-$\beta$ type; i.e., $V\left(x\right)=\frac{1}{2}x^{2}+
\frac{1}{4}x^{4}$. Both the mass $\mu$ and the lattice constant are set to be unit. The parameter $\gamma$ is tunable; it specifies the comparative strength of the NNN coupling. $\gamma=0$ corresponds to the conventional FPU-$\beta$ system.

We employ the reverse non-equilibrium molecular dynamics simulation method (RNEMD) \cite{RNEMD} to build the nonequilibrium stationary state across the system. Compared with the usual method that couples the system to two heat baths at its two ends, the RNEMD is advantageous in suppressing the boundary effects and therefore leads to a faster convergence to the stationary state. Meanwhile the RNEMD keeps the total energy and momentum of the system unchanged. Following the RNEMD \cite{RNEMD}, first the periodic boundary condition is imposed; then the chain (now a circle) is divided into $M$ slabs each contains $n=\frac{N}{M}$ particles. The instantaneous local kinetic temperature $T_{k}$ in slab $k$ ($k=1,\cdots,M$) is defined by $T_{k}\equiv \frac{1}{nk_{B}} \sum_{i=n(k-1)+1}^{nk}{p_{i}^{2}}$, where $k_{B}$ is the Boltzmann constant (set to be unit) and the sum goes over all $n$ particles in slab $k$. The temperature profile of the stationary state can be represented by the time average of $T_{k}$ ($k=1,\cdots,M$), denoted by $\langle T_{k}\rangle$. Next, slab $1$ is set to be the cold slab and slab $\frac{M}{2}+1$ as the hot slab. The heat flux is generated by exchanging the momenta of the hottest (coldest) particle in the cold (hot) slab at a fixed frequency $f_{\rm exc}$ (the exchange frequency). This procedure results in a redistribution of a certain amount of kinetic energy $\Delta E=\sum{\frac{1}{2} (p_{h}^{2}-
p_{c}^{2})}$ during time $t$. Here the subscript $h$ and $c$ refer to the hot and cold particles whose momenta are interchanged, and the sum takes all exchanges in time $t$. In the stationary state the relaxation of $\Delta E$ will drive two heat fluxes of $J= \frac{\Delta E}{2t}$ across the system, because due to the periodic boundary condition we have in fact two identical chains between the hot and cold slabs. $L\equiv \frac{N}{2}-n$ is the effective length of the two chains (deducting the hot and cold slabs) and $J$ is the heat flux crossing each of them. Once the stationary state is reached, a temperature gradient between the hot and cold slabs is expected and the thermal conductivity $\kappa$ can then be measured by assuming the Fourier law; i.e., $\kappa=-\frac{J}{\nabla T}$.

We start our simulations with a fully thermalized chain at temperature
$T=2.5$. The velocity-Verlet algorithm \cite{AllenLiquids} with
a time step $0.01$ is used to evolve the system, and $M=80$ and $f_{\rm exc}=0.1$ are adopted for the RNEMD. For each system size a transient stage of time $10^{6}$ is discarded (which has been verified to be long enough for reaching the stationary state); then the next evolution of time $10^{7}$ is performed for the time average. We have checked that our results do not qualitatively depend on the particular parameter values taken here.

\begin{figure}
\vskip-.2cm
\hskip-.4cm
\includegraphics[width=9.0cm]{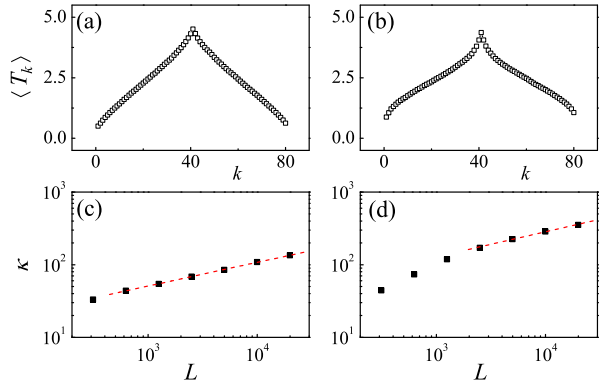}
\vskip-.4cm
\caption{(Color online) Top panels: the temperature profile for $\gamma=0$ (a) and $\gamma=1$ (b) with the effective system size $L=2496$. Bottom panels: the heat conductivity $\kappa$ versus $L$ for $\gamma=0$ (c) and $\gamma=1$ (d). The dashed lines are for the best fitting of $\kappa\sim L^\alpha$, suggesting $\alpha=0.325\pm0.002$ for $\gamma=0$ (c) and $\alpha=0.35\pm0.02$ for $\gamma=1$ (d).}
\end{figure}

Before presenting our main results, it is interesting to make a quick comparison between our simulations and those by different methods for $\gamma=0$ \cite{key-8} and $\gamma=1$ \cite{key-18}. For $\gamma=0$, i.e. the conventional FPU-$\beta$ system, Fig. 1(a) shows the temperature profile for $L=2496$; it can be seen that a constant temperature gradient is well established between the hot and cold slabs. In addition, for larger system sizes the temperature profiles (not shown) have been checked to be the same upon a rescaling.  Fig. 1(c) shows the dependence of $\kappa$ on the effective system size $L$; it suggests that $\kappa$ diverges as $\sim L^{\alpha}$ with $\alpha=0.325\pm0.002$ which we emphasize to be very close to the predicted value $\frac{1}{3}$ by the hydrodynamic theory \cite{key-14} and the result of a recent careful numerical study \cite{key-8}. For $\gamma=1$ we have obtained the similar results as $\gamma=0$ [see Fig. (b) and (d)], but the best fitting performed over $2496\le L\le 19968$ gives $\alpha=0.35\pm0.02$ instead [see Fig. 1(d)]. Note that our $\alpha$ value for $\gamma=1$ is remarkably different from that given in Ref. \cite{key-18} where $\alpha$ was evaluated over much shorter system sizes ($L<2000$) and therefore fails to capture the divergence of $\kappa$ in the thermodynamical limit.

\begin{figure}
\vskip-.2cm
\hskip-0.4cm
\includegraphics[width=7.5cm]{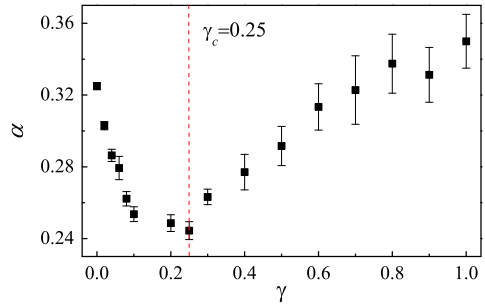}
\vskip-0.3cm
\caption{(Color online) The dependence of $\alpha$ on the parameter  $\gamma$. The vertical dashed line indicates $\gamma_{c}=0.25$. Error bars give the standard error for evaluating $\alpha$ by linearly fixing $\log \kappa$ versus $\log L$. }
\end{figure}

The first main result of our study is shown in Fig. 2 for the divergence exponent $\alpha$ versus the parameter $\gamma$. It can be seen that as $\gamma$ changes from $0$ to $1$, $\alpha$ decreases and reaches its minimum $\alpha_{\min}\approx 0.24$ at $\gamma\approx \gamma_{c}=0.25$, then increases up to about $0.35$ at $\gamma=1$ with a trend of saturation \cite{note}. The fact that $\alpha$ changes continuously in a range is in clear contrast to the existence of the general $\alpha$ value(s) independently of the dynamics.

As phonons are the heat energy carriers in our lattices, $\alpha$ takes its minimum at $\gamma_c$ implies that around $\gamma_c$ the NNN coupling may enhance the phonon scattering and thus give rise to smaller values of $\alpha$. To clarify this point the phonon dispersion relation turns out to be very suggestive. Keeping the harmonic coupling in both the NN and NNN interactions, the dispersion relation reads $\omega_{q}=2[\sin^{2} \frac{q}{2}+
\gamma\sin^{2}q]^{\frac{1}{2}}$, where $q$ is the wave number and $\omega_{q}$ the corresponding frequency. Interestingly, $\gamma_{c}$ is a transition value for the phonon dispersion relation as well (see Fig. 3): For $\gamma\leq\gamma_{c}$ the maximum frequency $\omega_{\pi}=2$ corresponds to the boundary of the Brillouin zone $q=\pi$ but for $\gamma>\gamma_{c}$ it grows
larger than $\omega_{\pi}$ and moves away from the boundary. For
$\gamma=\gamma_{c}$ the group velocity is close to zero in a wider $q$ region. This property favors the DBs in the presence of the nonlinearities \cite {OurNext} which we will discuss later [see Fig. 5(b)].

\begin{figure}
\vskip-.2cm
\hskip-0.6cm
\includegraphics[width=7.5cm]{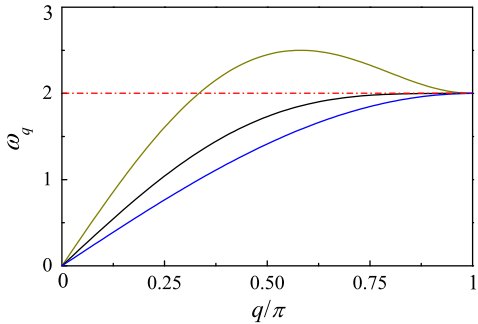}
\vskip-0.3cm
\caption{(Color online) Phonon dispersion relation for the FPU-$\beta$ system with the NNN interactions. The curves from bottom to top correspond to $\gamma=0,$ $0.25$, and $1$ respectively. The horizontal dash-dotted line indicates $\omega_{\pi}=2$. }
\end{figure}

Since the work by Peyrard \cite{PeyrardThermalDBs} it has been known that the temperature activated DBs may be crucial for the energy transport and other dynamical processes. Interesting examples include the melting transitions in solids and folding in polypeptide chains \cite{DBPhysToday}. To study if the DBs may have any effects on the heat conduction in our system, it is necessary to check if DBs exist at the focused temperature $T=2.5$. We apply the method in Ref. \cite{fai4DBs}; i.e., a chain of $N=2000$ particles is first thermalized at temperature $T=2.5$, then the heat baths are removed and the absorbing boundary conditions are imposed to the two ends of the chain \cite{key-38}. After all the mobile excitations such as phonons and solitary waves are absorbed, a few standing breathers may emerge in the internal segment of the chain. The snapshot of the energy profile after a long time ($8\times 10^5$) absorbtion is presented in Fig. 4(a) and (b) for $\gamma=0$ and $\gamma=\gamma_{c}$, respectively; in both of them the DBs can be well recognized. We have verified that this is also the case for other $\gamma$ values in $[0,1]$.

Now we study the interactions between the DBs and the phonons. For this aim we calculate the power spectra $P(\omega)$ of the residual
thermal fluctuations after long time absorbing. To facilitate the computation short chains of size $N=200$ are considered. The results for $\gamma=0$ and $\gamma=\gamma_{c}$ are plotted in Fig. 4(c) and Fig. 4(d) for a comparison; It shows that for $\gamma=0$ the DBs' frequencies are outside the linear phonon band of $0 \le \omega \le \omega_\pi$, in agreement with the classical DB theory \cite{SieversDBs}. Due to this frequency mismatch we may conjecture the lack of interactions between the DBs and the phonons, which in turn implies the null effects of the DBs on the heat conduction. However, in clear contrast, for $\gamma=\gamma_{c}$ a portion of the DBs' frequencies appear inside the linear phonon band, suggesting the existence of the in-band DBs \cite{key-1,key-2,key-3}. This is well shown in the insert of Fig. 4(d) where the collective modes in the linear phonon band can be clearly recognized. As the in-band DBs can randomly distribute along the lattice and interact with phonons, they introduce an inherent disorder \cite{SieversDBs} and play roles of random scatters to the phonons. This may explain why $\alpha$ is smaller in the case of $\gamma=\gamma_{c}$. If this is true, combining the results given in Fig. 2 we may expect that as $\gamma$ is increased, the interactions between the in-band DBs and the phonons would become stronger and stronger (weaker and weaker) for $0\le \gamma\le \gamma_c$ ($\gamma_c<\gamma\le 1$). To check if this is the case, we define $\varepsilon={\int_{0}^{\omega_{\pi}} P(\omega)d\omega} /{\int_{0}^{\infty}P(\omega)d\omega}$, the ratio of the energy of the collective modes within the linear phonon band to the total energy of the residual thermal fluctuations, as a measure of the interaction intensity between the in-band DBs and the phonons. For several typical $\gamma$ values we first calculate the power spectra in the same way as in Fig. 4(c) and (d), then evaluate $\varepsilon$ and summarize the results in Fig. 5(a). It shows that $\varepsilon$ versus $\gamma$ is in good consistence with $\alpha$ versus $\gamma$ (see Fig. 2). This consistence verifies that the heat conduction behavior is indeed dependent on the DB-phonon interactions in our system. (Note that in Fig. 5(a) the maximum of $\varepsilon$ does not correspond to $\gamma_c$ exactly but a slightly smaller $\gamma$ value; this discrepancy may be a result of the big statistical errors in the evaluated power spectra of the residual thermal fluctuations where short chains of $N=200$ have to be used on account of computation cost.)

\begin{figure}
\vskip-.2cm
\hskip-0.4cm
\includegraphics[width=9cm]{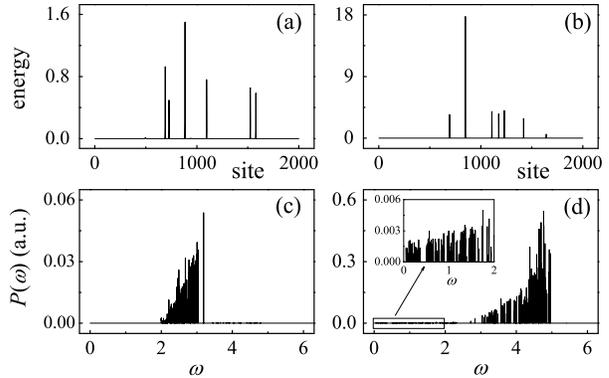}
\vskip-0.4cm
\caption{(a) and (c): The snapshot of the energy distribution
and the power spectrum of the residual thermal fluctuations for $\gamma=0$. (b) and (d): the corresponding results for $\gamma=0.25$. While for $\gamma=0$ no in-band components can be identified, there are considerable in-band components for $\gamma=0.25$ [see the insert for a zoom of the boxed in-band region in (d)]. }
\end{figure}

In the following let us turn to the question why the DB-phonon interactions are the strongest at $\gamma=\gamma_c$. Our study [see Fig. 5(b) later] suggests the DBs' concentration is the highest at $\gamma=\gamma_c$, which we conjecture results in the highest  probability for the DBs to interact with the phonons. The concentration of DBs is approximately $e^{-\frac{e_{\rm sh}}{k_{B}T}}$ in an equilibrium state, where $e_{\rm sh}$ is the energy threshold for creating the DBs \cite{SieversDBs}. It implies that $e_{\rm sh}$ is the smallest if the DBs' concentration is the highest (at $\gamma=\gamma_c$). To check if this is the case we consider an energy relaxation process. Initially we give a kinetic energy $e(0)$ to the particle centered at the chain of $N=2000$, a zero velocity to all others and a zero displacement to all particles. Then absorbing boundary conditions are imposed, and after a transient time of $2\times10^{5}$ we calculate $e(t)$, the total energy remained in the chain at time $t$, up to $t=8\times10^{5}$. We find it decays as $e(t)\sim t^{-\beta}$ \cite{OurNext} and the exponent $\beta$ depends on both the initial excitation energy $e(0)$ and $\gamma$: When $e(0)$ is small enough harmonic behaviors manifest themselves and $\beta\rightarrow 0.5$ as being pointed out in \cite{EnergyDecay}; but when $e(0)$ is large enough long-lived DBs could form and thereby $\beta\rightarrow 0$. We increase $e(0)$ progressively and find it is exactly at $\gamma=\gamma_c$ where the signal for the DBs appear first [see Fig. 5(b)]. Hence we may conclude for $\gamma=\gamma_c$ the value of $e_{\rm sh}$ is the smallest and the concentration of the DBs is the highest.

\begin{figure}
\includegraphics[height=6cm]{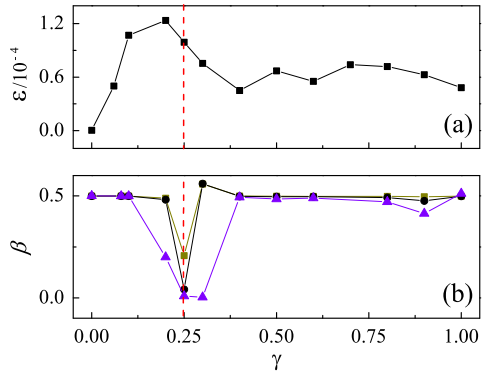}
\vskip-0.4cm
\caption{(Color online) (a) $\varepsilon$ versus $\gamma$ and (b) $\beta$ versus $\gamma$ for $e(0)=0.005$ (squares), 0.045 (dots) and 0.18 (triangles). The vertical dashed line indicates $\gamma_{c}=0.25$ in both panels.}
\end{figure}

In summary, we have studied a one dimensional lattice of the FPU-$\beta$ type with the NNN coupling. We have shown that tuning the NNN coupling may change the concentration of the DBs, which in turn vary the DB-phonon interaction intensity, and consequently affect the heat conduction characteristics of the system. Depending on the NNN coupling, the divergence exponent $\alpha$ of the heat conductivity may continuously change from 0.24 to about $\frac{1}{3}$. In contrast to the previously suggested generality class(es), our study reveals a new regime in one dimensional heat conduction featuring the interactions between nonlinear excitations and phonons. At present the understanding to this new regime is still primitive; further investigations are necessary and desired.

\begin{acknowledgments}
This work is supported by the NNSF of China (Grants No. 10805036, No. 10925525, No. 10975115), the RFDP of China (Grant No. 20100121110021), and the start-up fund (No. 022390) from Fuzhou University of China.
\end{acknowledgments}

\end{CJK}

\begin{thebibliography}{References}
\bibitem{note} This is qualitatively different from the result in Ref. \cite{key-18} where much shorter chains ($L<2000$) were considered and $\alpha$ was reported to increase with $\gamma$ monotonically.
\end{thebibliography}
\end{document}